\documentstyle[11pt,newpasp,twoside,epsf]{article}
\def\lsim{\lower.5ex\hbox{$\; \buildrel < \over \sim \;$}}
\def\gsim{\lower.5ex\hbox{$\; \buildrel > \over \sim \;$}}

\def\edcomment#1{\iffalse\marginpar{\raggedright\sl#1\/}\else\relax\fi}
\reversemarginpar

\begin{document}
\title{LiBeB Evolution: Three Models}
\author{Reuven Ramaty}
\affil{NASA/GSFC, Greenbelt, MD 20771, USA}
\author{Richard E. Lingenfelter}
\affil{CASS/UCSD, LaJolla, CA 92093, USA}
\author{Benzion Kozlovsky}
\affil{Tel Aviv University, Israel}

\begin{abstract}

We consider the three principal LiBeB evolutionary models, CRI in
which the cosmic-ray source at all epochs of Galactic evolution is
the average ISM, CRI+LECR in which metal enriched low energy cosmic
rays (LECRs) are superimposed onto the CRI cosmic rays, and CRS in
which the cosmic-ray source, accelerated in superbubbles, is
constant, independent of the ISM metallicity. By considering the
evolutionary trend of log(Be/H) vs. both [Fe/H] and [O/H], we
demonstrate that the CRI model is energetically untenable. We
present evolutionary trends for $^{11}$B/$^{10}$B and B/Be which,
combined with future precision measurements, could distinguish
between the CRS and CRI+LECR models. We show that delayed LiBeB
synthesis in the CRS model, due to the transport of the cosmic rays,
could explain why log(Be/H) is steeper vs. [O/H] than vs. [Fe/H]. We
also show that delayed deposition of Fe into star forming regions,
due to its incorporation into high velocity dust, could provide an
explanation for the possible rise of [O/Fe] with decreasing [Fe/H].
Observations of refractory and volatile $\alpha$-elements could test
this scenario. There seems to be a need for pregalactic or
extragalactic $^6$Li sources. \end{abstract}

\section{Introduction}

Cosmic-ray driven nucleosynthesis has been known to be important
for the origin of the light elements Li, Be and B (LiBeB) for
three decades (Reeves, Fowler, \& Hoyle 1970). But constraints on
the relevant evolutionary models could only be obtained after
LiBeB abundances of low metallicity stars started to become
available (e.g. Ryan et al. 1990). The principal three models
currently considered are: (i) the cosmic-ray interstellar model
(hereafter CRI), in which the cosmic-ray source composition at all
epochs of Galactic evolution is assumed to be similar to that of
the average ISM at that epoch (Vangioni-Flam et al. 1990; Fields
\& Olive 1999); (ii) the CRI+LECR model, in which metal enriched
low energy cosmic rays (LECRs) are superimposed onto the CRI
cosmic rays (Cass\'e, Lehoucq, \& Vangioni-Flam 1995;
Vangioni-Flam et al. 1996; Ramaty, Kozlovsky, \& Lingenfelter
1996); and (iii) the cosmic-ray superbubble model (hereafter CRS),
in which the cosmic-ray source composition is taken to be
constant, independent of the ISM metallicity (Ramaty et al.
1997;2000; Higdon, Lingenfelter, Ramaty 1998). Both the CRS cosmic
rays and the LECRs are thought to be accelerated out of supernova
enriched matter in superbubbles.

Because of the excess of the observed Be abundances in low
metallicity stars over the predictions of the CRI model, and
motivated by reports of the detection of C and O nuclear gamma-ray
lines from the Orion star formation region (Bloemen et al. 1994),
LECRs, enriched in C and O relative to protons and $\alpha$
particles, were superimposed on the CRI cosmic rays, hence the
CRI+LECR model . These LECRs, with  maximum energies not exceeding
about 100 MeV/nucleon, were thought (e.g. Ramaty 1996) to be
responsible for the gamma rays reported from Orion. It was suggested
that such enriched LECRs might be accelerated out of metal-rich
winds of massive stars and supernova ejecta (Bykov \& Bloemen 1994;
Ramaty et al. 1996; Parizot, Cass\'e, \& Vangioni-Flam 1997) by an
ensemble of shocks in superbubbles (Bykov \& Fleishman 1992; Parizot
Cass\'e, \& Vangioni-Flam 1997). The Orion gamma-ray data, however,
have been retracted (Bloemen et al. 1999). Nonetheless, as the
possible existence of the postulated LECRs remains, new gamma-ray
line data are needed to determine the role of LECRs in LiBeB
production.

Recent O abundance data, which suggest that [O/Fe] increases with
decreasing [Fe/H] at low metallicities (Israelian et al. 1998;
Boesgaard et al. 1999), led Fields and Olive (1999) to reexamine the
viability of the CRI model. More recent measurements (Fulbright \&
Kraft 1999; Westin et al. 1999) argue against such an [O/Fe]
increase. But as demonstrated in (Ramaty et al. 2000), and also
shown below, this model is inconsistent with cosmic-ray energetics,
an [O/Fe] increase notwithstanding.

Alternatively, it was suggested (Lingenfelter, Ramaty, \&
Kozlovsky 1998; Higdon et al. 1998), that the Be evolution can be
best understood in the CRS model, in which the cosmic-ray metals
at all epochs of Galactic evolution are accelerated predominantly
out of supernova ejecta. Lingenfelter et al. (1998) and
Lingenfelter \& Ramaty (1999) showed that the arguments (e.g.
Meyer, Drury, \& Ellison 1997) against the supernova ejecta origin
of the current epoch cosmic rays can be answered, and Higdon et
al. (1998) and Higdon, Lingenfelter, \& Ramaty (1999) showed that
the most likely scenario is collective acceleration by successive
supernova shocks of ejecta-enriched matter in the interiors of
superbubbles. This scenario is consistent with the delay between
nucleosynthesis and acceleration (time scales $\sim$10$^5$ yr),
suggested by the $^{59}$Co and $^{59}$Ni observations (Wiedenbeck
et al. 1999). In both the CRS and CRI+LECR models, the bulk of the
Be in the early Galaxy is produced by accelerated C and O
interacting with ambient H and He. That these ``inverse reactions"
are dominant in the early Galaxy was first suggested by Duncan,
Lambert, \& Lemke (1992).

In the present paper we present results from a complete set of
LiBeB evolutionary calculations for all three models using our
production code described in Ramaty et al. (1997) and evolutionary
code detailed in Ramaty et al. (2000).

\begin{figure} \plottwo{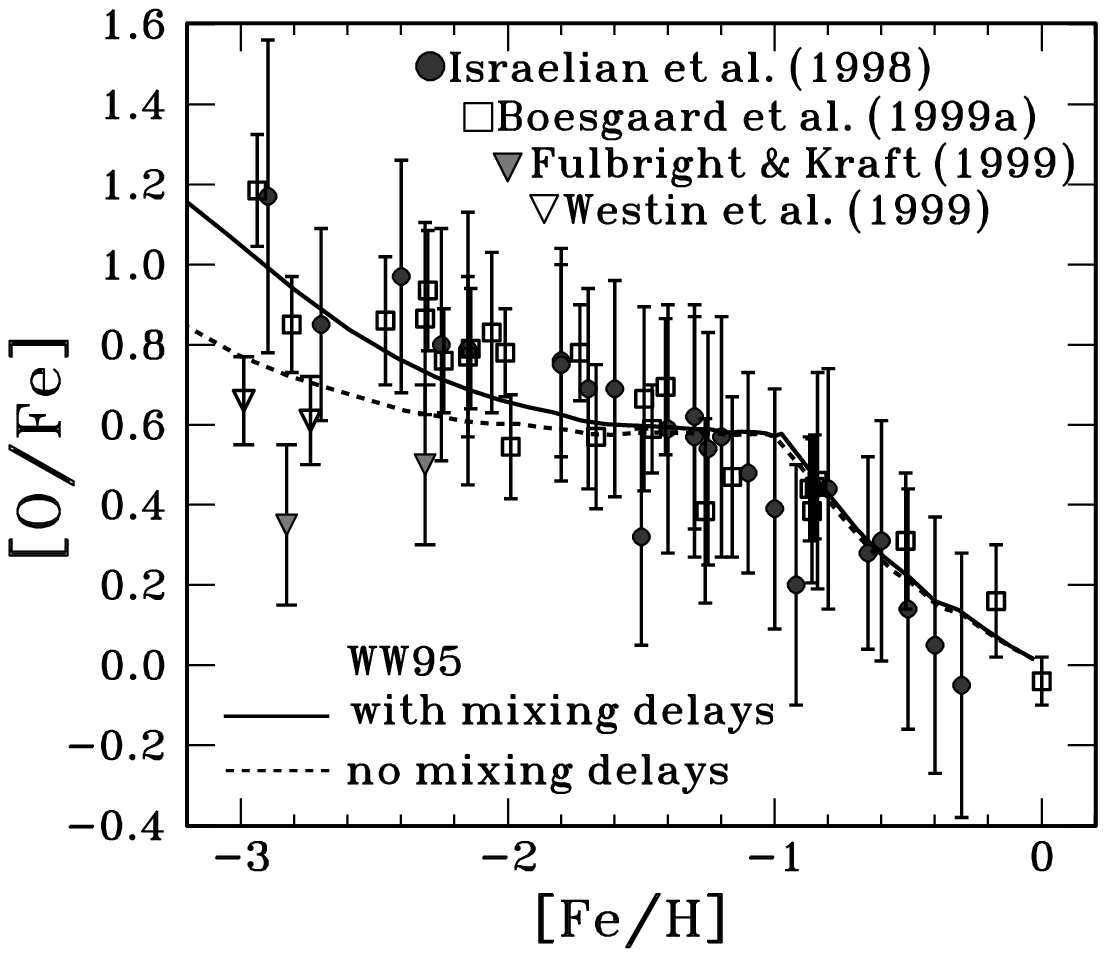}{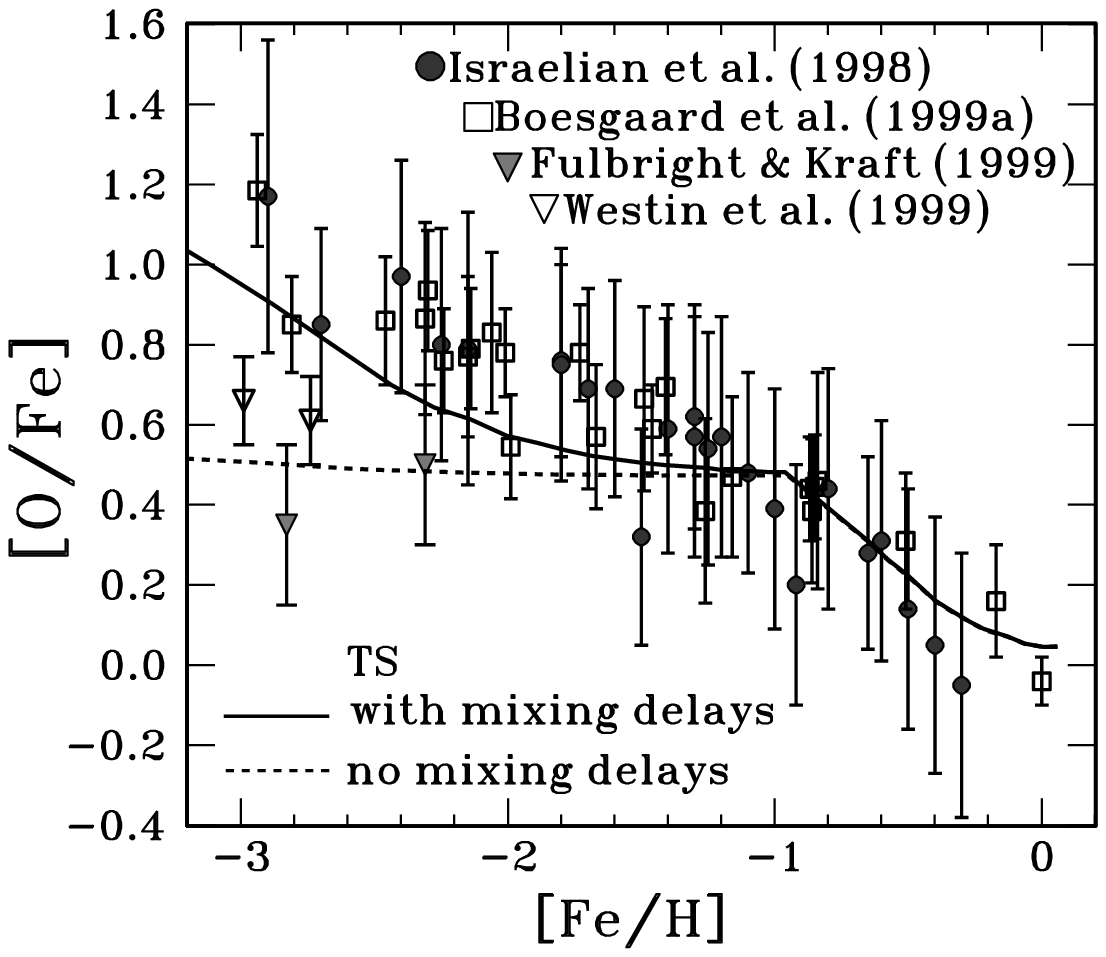} \caption
{Galactic evolution of [O/Fe] vs. [Fe/H] for supernova O and Fe
yields taken from minimum Fe yield models of Woosley \& Weaver (1995,
WW95 shown in the left panel) and Tsujimoto \& Shigeyama (1998, TS
shown in the right panel), assuming zero and finite Fe and O mixing
delay times, $\tau_{\rm Fe}$(mix) =$30$Myr and $\tau_{\rm O}$(mix)
=$1$Myr, and including contributions from Type Ia supernovae. The
characteristic halo and disk infall times are $10 {\rm Myr}$ and $5
{\rm Gyr}$, respectively, the star formation rate coefficient is $0.5
{\rm Gyr}^{-1}$, and the ratio of halo-to-disk masses is 0.1. (See
Ramaty et al. 2000 for more details).} \end{figure}

\section{Analysis}

Figure~1 shows the evolution of [O/Fe] as a function of [Fe/H]. In
order to account for the possible rise of [O/Fe] with decreasing
[Fe/H], we introduced mixing delays, i. e. the delayed deposition
of the synthesized products into the star forming regions due to
differences in transport and mixing. We choose a short mixing time
for oxygen because we expect the bulk of the O and other volatiles
in the ejecta to mix with the ISM after the remnant slows down to
local sound speeds. But we consider longer mixing times for Fe,
assuming that the bulk of the ejected Fe is incorporated into high
velocity refractory dust grains which continue moving for longer
periods of time before they stop and can be incorporated into
newly forming stars. The incorporation of a large fraction of the
synthesized Fe into dust grains is supported by observations of
both supernova 1987A and the Galactic 1.809 MeV gamma-ray line
resulting from the decay of $^{26}$Al (for more details and
references see Ramaty et al. 2000). We see that with the mixing
delays (solid curves in Figure~1) both the WW95 and TS cases (see
figure caption) become consistent with the Israelian et al. (1998)
and Boesgaard et al. (1999a) data, showing that delayed Fe
deposition could indeed be the cause for the rise of [O/Fe]. In
this connection, it is interesting to note that, unlike [O/Fe],
the abundance ratios of the $\alpha$-nuclei Mg, Si, Ca and Ti
relative to Fe do not increase with decreasing [Fe/H] below [Fe/H]
$= -1$ (Ryan, Norris, \& Beers 1996). This may be consistent with
the fact that these elements are also refractory, and thus are
affected by mixing in the same way as is Fe. A test may be
provided by sulfur, which is volatile, and thus should show a rise
similar to the rise of [O/Fe] vs. decreasing [Fe/H] (G. Israelian,
private communication, 1999).

\begin{figure}
\plottwo{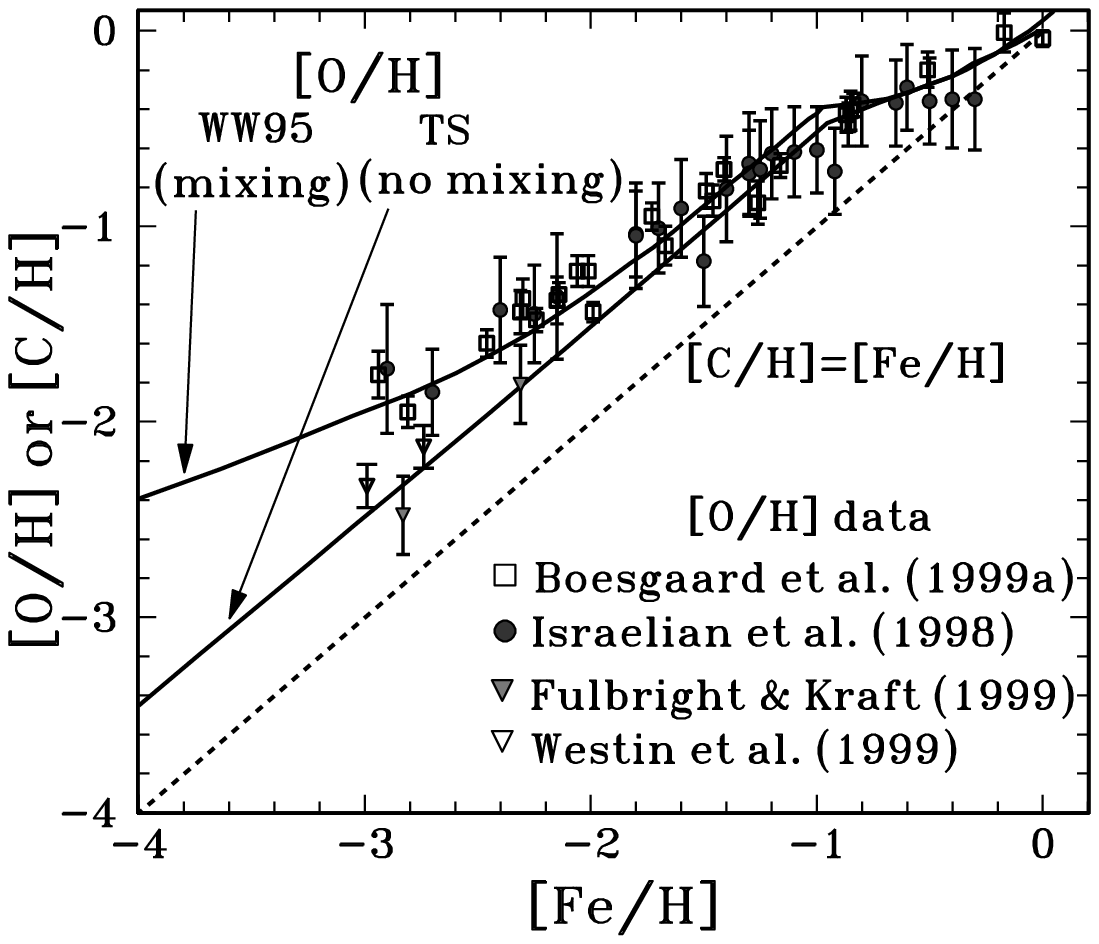}{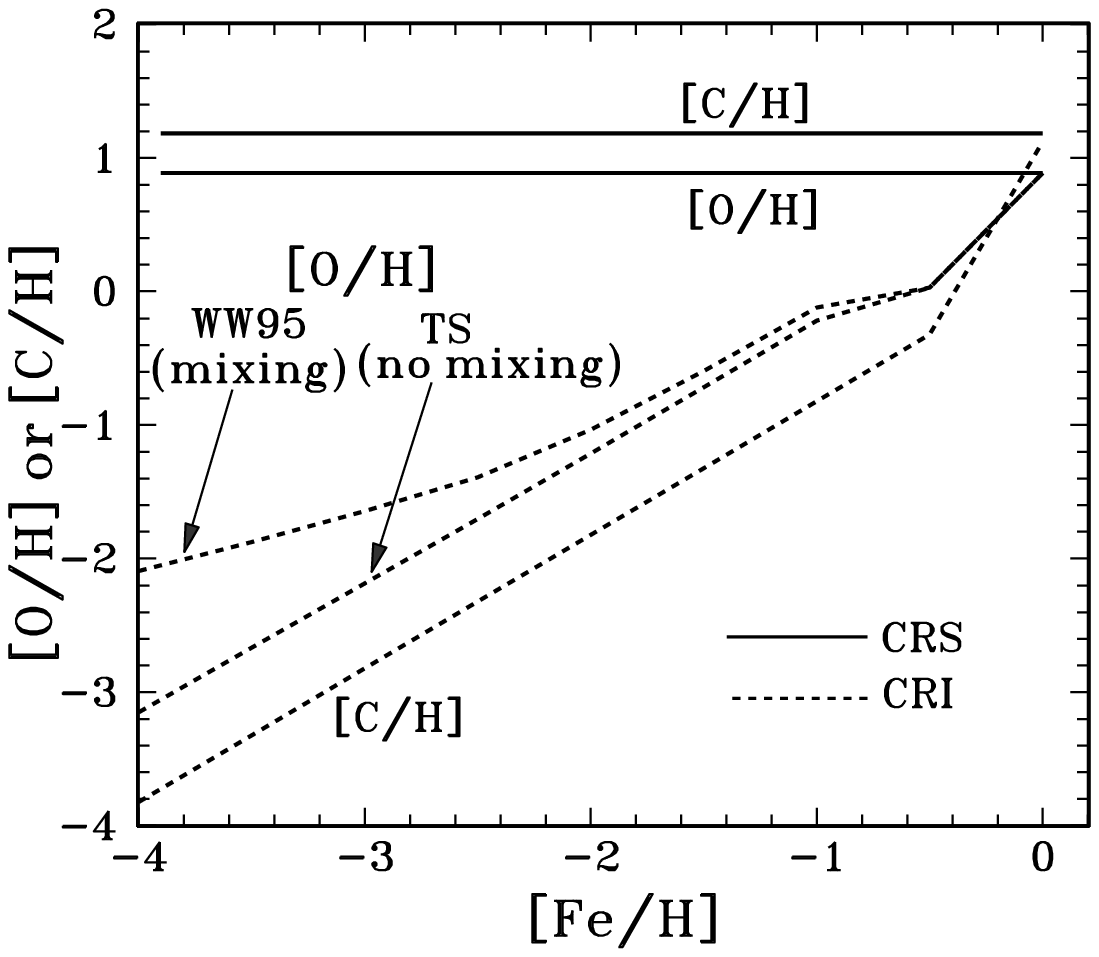} \caption{The employed ISM
(left panel) and cosmic-ray source (right panel) C and O
abundances as functions of [Fe/H]. For the ISM, [O/H] is obtained
from the evolutionary calculation for the two limiting cases, WW95
with mixing delays and TS with no mixing delays (Figure~1). [C/Fe]
is taken constant. For the cosmic rays the abundances are
normalized to the solar values, i.e.
[X/Fe]$\equiv$log(X/Fe)$-$log(X/Fe)$_\odot$ where X stands for C
or O. For the CRS model [C/H] and [O/H] are independent of [Fe/H].
For the CRI model, [C/H] and [O/H] are based on the ISM values
with enhancement factors consistent with shock acceleration theory
(see text), except at [Fe/H]=0 where the cosmic ray CRS and CRI
values are equal.}
\end{figure}

Figure~2 shows the employed C and O abundances of the ISM and the
cosmic ray source. For the ISM (left panel) we take
[C/H]$\simeq$[Fe/H], because at early times both the C and Fe come
primarily from core collapse supernovae of massive stars, while at
later times the increased C contribution from the winds of
intermediate mass stars is compensated by the Fe contribution from
the thermonuclear supernovae of the white dwarf remnants of such
stars (see Timmes, Woosley \& Weaver 1995). The O abundances follows
from the results of evolutionary calculations shown in Figure~1.
Unlike in Ramaty et al. (2000), where we used a constant He
abundances, here we allow He/H (by number) to vary slowly from 0.08
at very low metallicities to 0.1 at [Fe/H]=0. For the cosmic-ray
source (right panel), we define the logarithmic ratios [C/H] and
[O/H] in the same way as is done for the corresponding ISM values,
including normalization to solar (not current epoch cosmic ray
source) abundances. As the CRS cosmic rays are accelerated primarily
out of supernova ejecta enriched superbubbles, [C/H] and [O/H] are
constant, set equal to current epoch cosmic-ray values. For the CRI
model we scale [C/H] and [O/H] to the ISM values with enhancement
factors of 1.5 and 2, consistent with the mass-to-charge dependent
acceleration of volatiles (Ellison, Drury \& Meyer 1997), except at
[Fe/H]=0 where the cosmic ray CRS and CRI values are equal. For
details on the rest of the employed cosmic-ray sources abundances
see Ramaty et al. (2000). For the LECRs we adopt the CRS abundances.
The CRS, CRI and LECR source energy spectra are power laws in
momentum with high energy exponential cutoffs (characteristic energy
$E_0$), which we set to an ultrarelativistic value for the CRS and
CRI cosmic rays and to 30 MeV/nucleon for the LECRs.

\begin{figure}
\plottwo{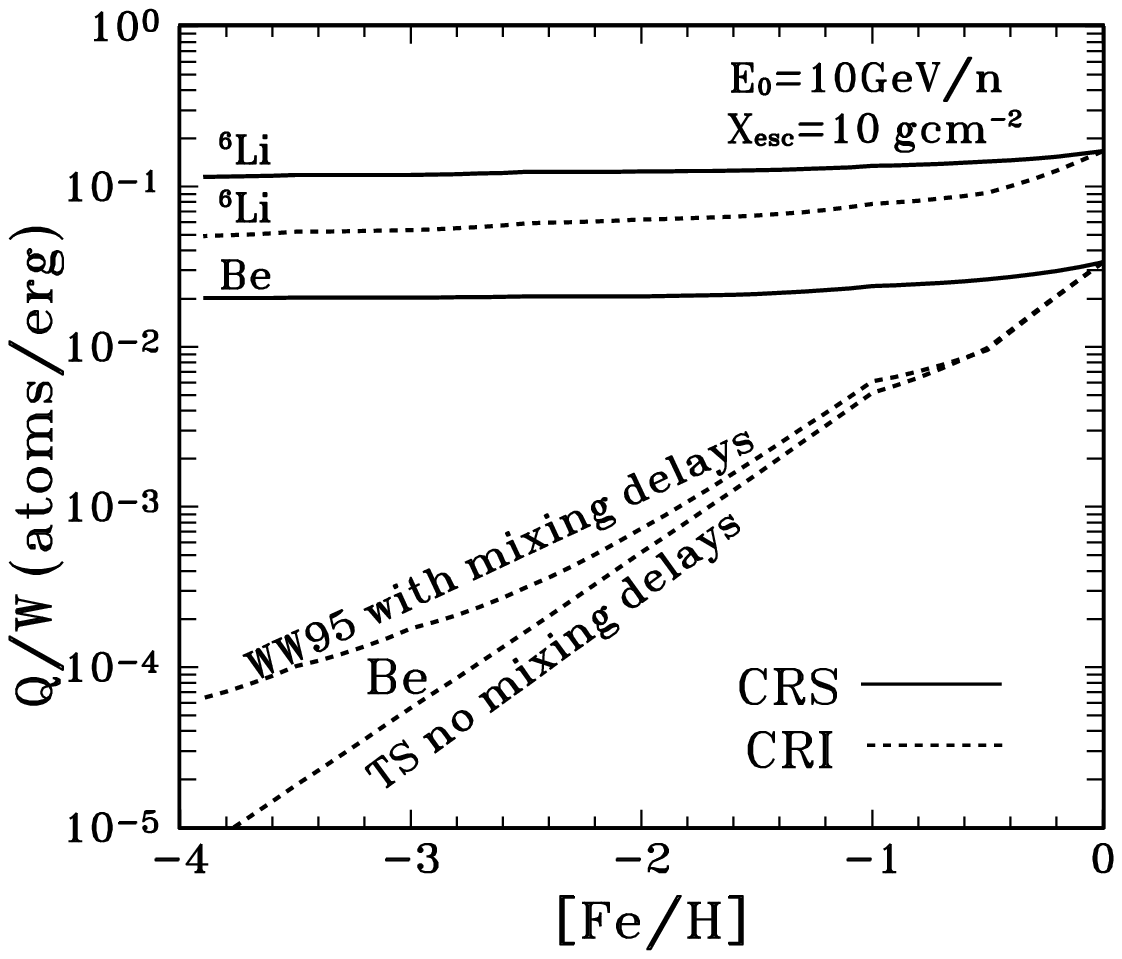}{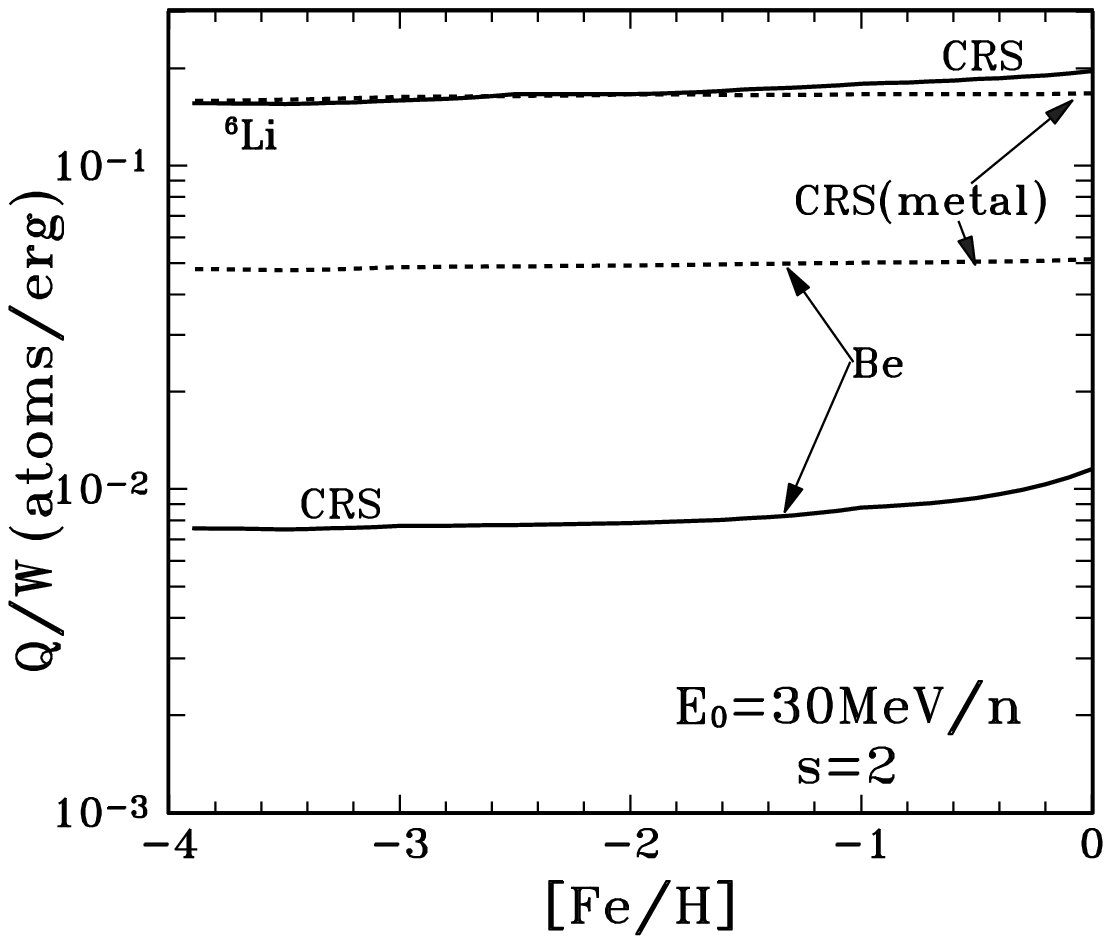} \caption{Number of Be and
$^6$Li atoms produced per unit cosmic-ray energy. Left panel --
for the CRS and CRI models using a cosmic-ray source momentum
power law spectral index of 2.5. For the CRI Be the two limiting
mixing cases are shown. For $^6$Li in both the CRS and CRI models,
and for Be in the CRS model, the mixing cases do not affect Q/W.
Right panel -- for the LECR model with a turnover energy of 30
MeV/nucleon, spectral index of 2, CRS and CRS(metal) abundances,
where the latter is identical to CRS, except that the proton and
$\alpha$ particle abundances are set to zero. For these abundances
the mixing cases have no effect. }
\end{figure}

Figure~3 shows the resultant $Q/W$'s, the total number of nuclei
$Q$ produced by an accelerated particle distribution normalized to
the integral cosmic-ray energy $W$, for a given source energy
spectrum, and cosmic ray and ambient medium compositions as
described above. We make the reasonable assumption that the
accelerated particle source energy spectrum is independent of
[Fe/H]. The resultant CRS and CRI $Q/W$'s for Be and $^6$Li (left
panel) are for ${\rm X}_{\rm esc} = 10~{\rm g cm}^{-2}$, typical
of currently inferred values for ``leaky box" cosmic-ray
propagation models. While $\alpha$$\alpha$ dominated  $Q(^6{\rm
Li})/W$ is not very different for the CRS and CRI models, $Q({\rm
Be})/W$ is drastically different for the two models, reflecting
the fact that efficient Be production in the early Galaxy can only
result from C and O enriched accelerated particles. The different
O abundances employed in the calculation of $Q/W$ for the two
mixing delays cases (see Figure~2) lead to significantly different
$Q({\rm Be})/W$'s for CRI model. The LECR $Q/W$'s (right panel)
show that, while removal of the protons and $\alpha$ particles
(the CRS(metal) composition) significantly increases $Q/W$ for Be,
it essentially leaves $Q(^6{\rm Li})/W$ unchanged, because the
lack of $^6$Li production by $\alpha$ particles is compensated by
a smaller W due to the absence of the protons and alphas.

\begin{figure}
\plottwo{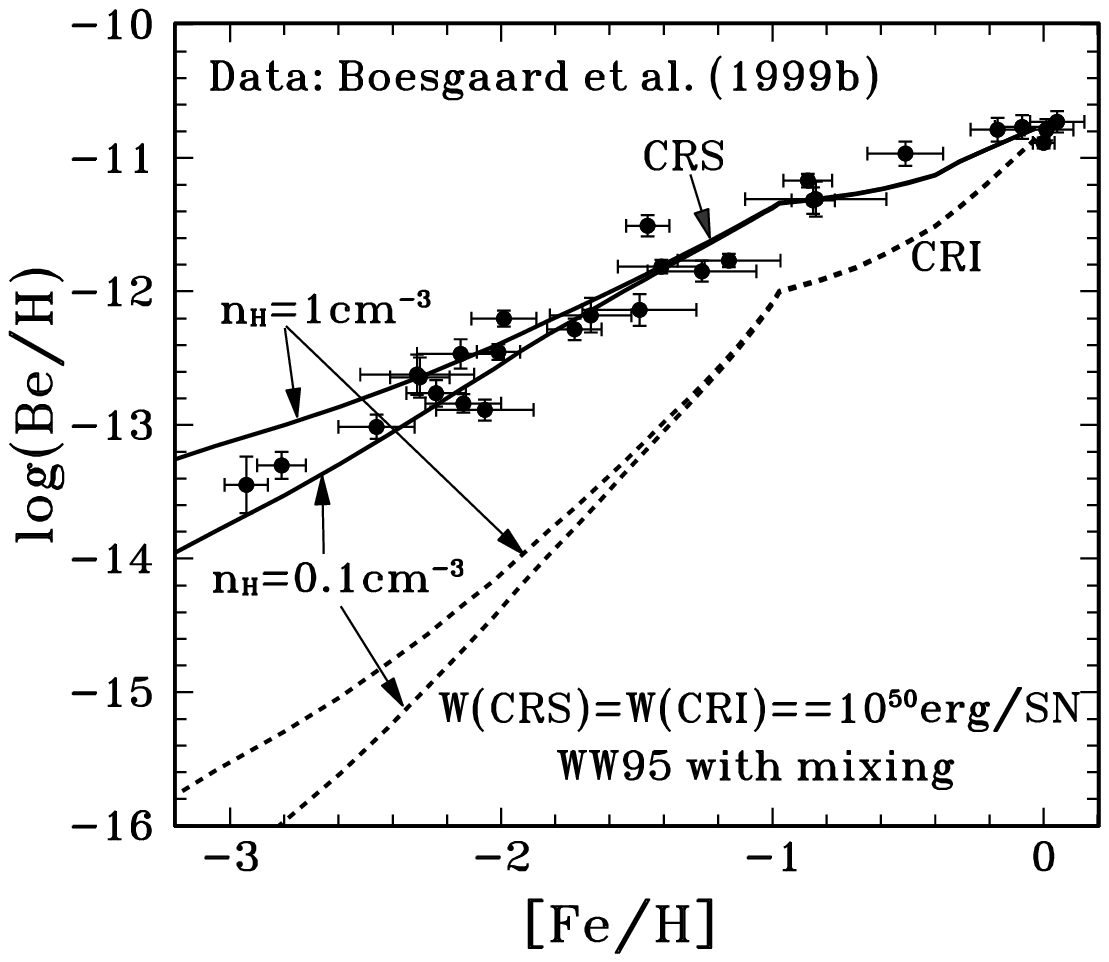}{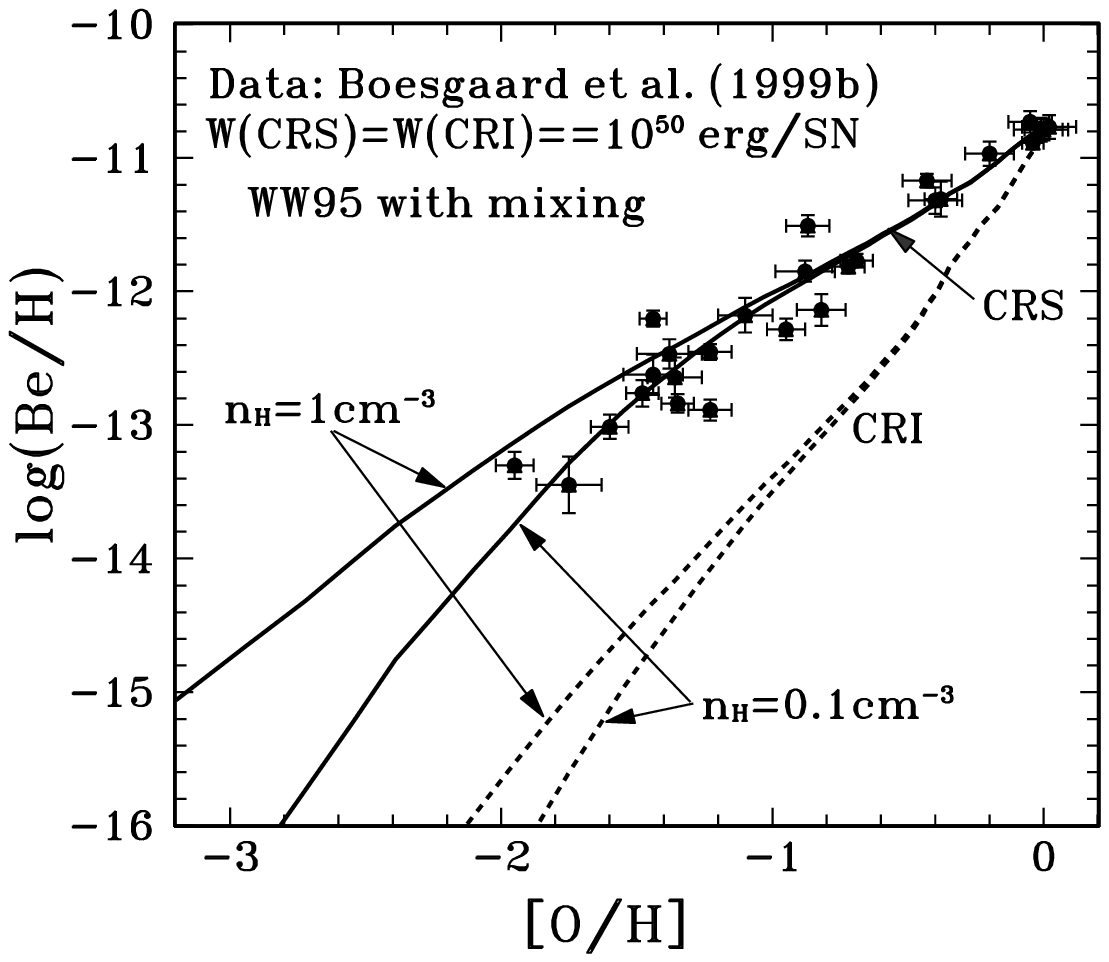} \caption{Be abundance evolution
for the CRS and CRI models, as a function of [Fe/H] (left panel)
and [O/H] (right panel). n$_{\rm H}$ is the density of ambient
hydrogen which influences the propagation of the cosmic rays and
determines the delay between the supernova explosion and the
deposition of the synthesized Be. For both the CRS and CRI model,
10$^{50}$ ergs per supernova are imparted to the cosmic rays.
Results for only WW95 with mixing delays are shown.}
\end{figure}

Figure~4 shows the Be evolution for the CRS and CRI models. In the
calculations, 10$^{50}$ erg per supernova are imparted to the cosmic
rays, a value in very good agreement with current epoch cosmic-ray
energetics. We note that even though the overall slope of log(Be/H)
vs. [Fe/H] is practically unity, while that of log(Be/H) vs. [O/H]
is significantly steeper (0.96$\pm$0.04 and 1.45$\pm$0.04,
respectively, Boesgaard et al. 1999b), the CRS model provides a good
fit to these evolutionary trends, particularly if n$_{\rm H}$ is
near 0.1. Such a low value might not be unreasonable for an average
halo hydrogen density if 10$^{10}$M$_\odot$ are spread over a few
kpc$^3$. The calculated log(Be/H) vs. [Fe/H] is flatter than
log(Be/H) vs. [O/H] because the delayed deposition of the
synthesized Be, caused by the low n$_{\rm H}$, is compensated by the
delayed Fe deposition, due to the incorporation of Fe in high
velocity dust, but not compensated by the very short delay of the
deposition of O, which is mostly volatile. As in Ramaty et al.
(2000), we see that the CRI model, normalized to a reasonable energy
in cosmic rays per supernova, severely underproduces the measured Be
abundances. However, unlike in that paper where we showed the result
only for log(Be/H) vs. [Fe/H], here we show that the same result
also holds for log(Be/H) vs. [O/H]. This removes the remaining
ambiguity concerning our argument against the result of Fields \&
Olive (1999), who claimed that the CRI model would be viable if
instead of Fe ejecta per supernova based on calculations, which are
somewhat uncertain, they used values based on their fit to
the increasing [O/Fe] with decreasing [Fe/H]. As there is no such
uncertainty concerning the O ejected masses, our present result
unequivocally demonstrates that the CRI model is untenable.

\begin{figure}
\plottwo{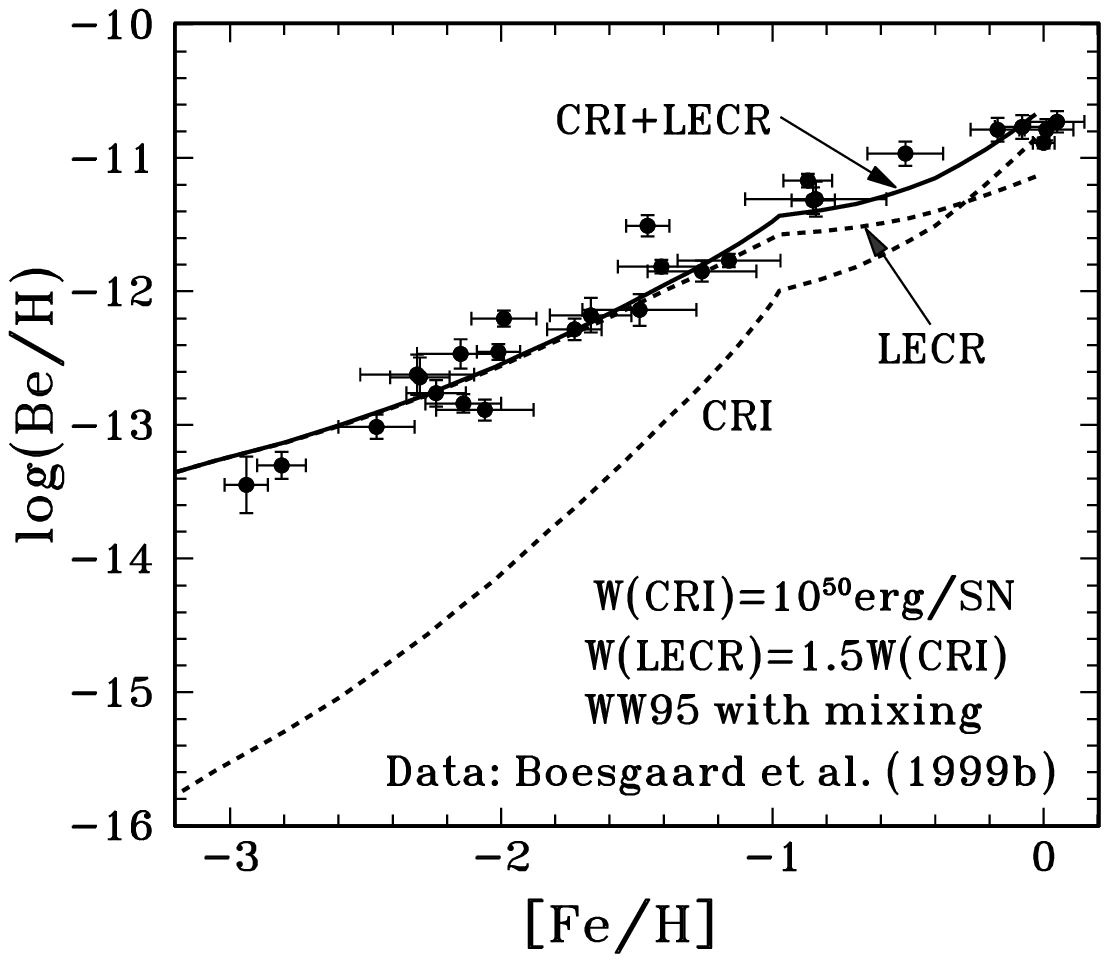}{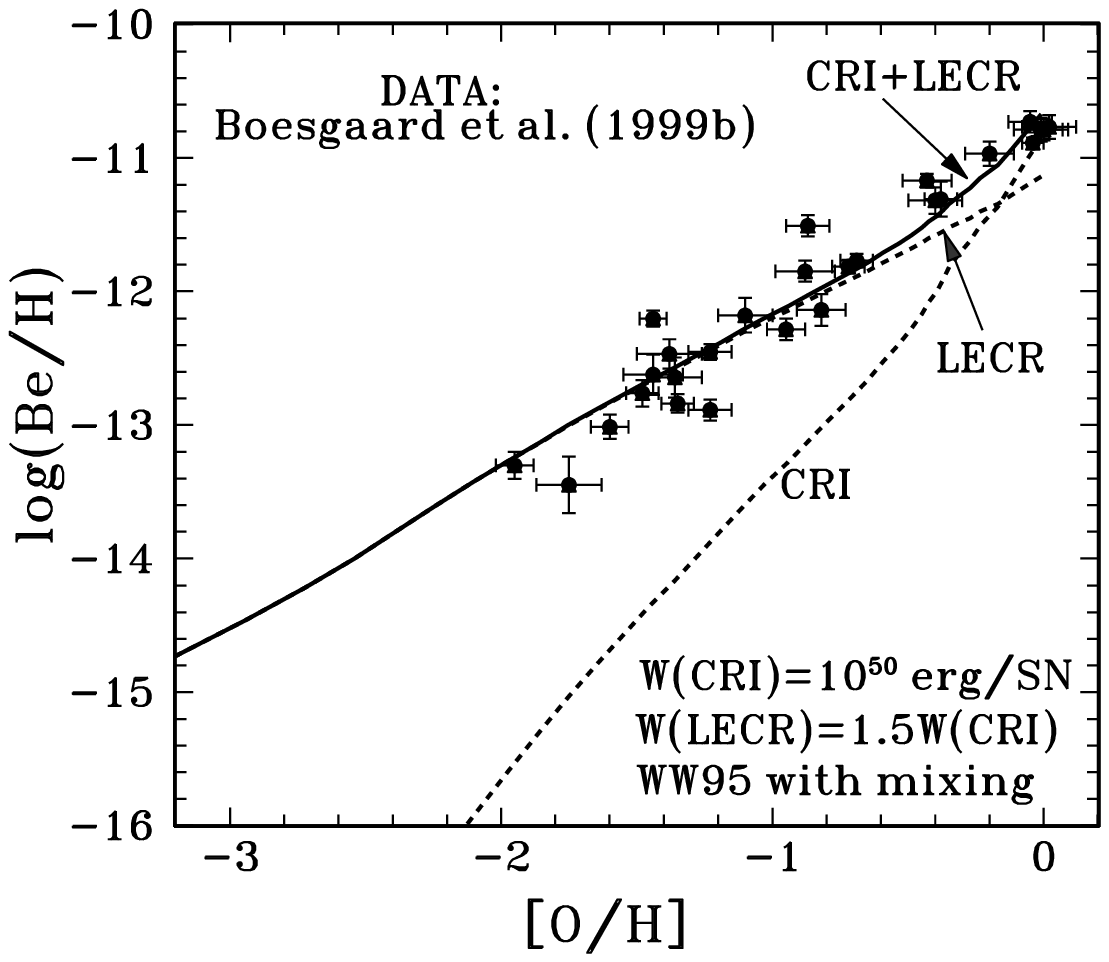} \caption{ Be abundance
evolution for the CRI+LECR model with a turnover energy of 30
MeV/nucleon, as a function of [Fe/H] (left panel) and [O/H] (right
panel). LECRs slow down much faster than higher energy
cosmic rays, hence there is no delay between the supernova
explosion and the deposition of the Be. 10$^{50}$ ergs
per supernova are imparted to the CRI cosmic rays and 50\% more to
the LECRs.}
\end{figure}

Figure~5 shows the Be evolution for the CRI+LECR model. Here, as
before, 10$^{50}$ erg per supernova are imparted to the CRI cosmic
rays, but in order to achieve a fit to the log(Be/H) vs. [Fe/H]
data, we had to add more energy (1.5$\times$10$^{50}$ erg per
supernova) in LECRs. The need for this increased energy can be seen
in Figure~3, where the $Q({\rm Be})/W$ for the LECR model with the
CRS composition (right panel) is lower by about a factor of 2 than
the corresponding value for the CRS model (left panel). Returning to
Figure~5, we see that the CRI+LECR model leads to a log(Be/H) vs.
[O/H] evolutionary curve with a slope which is only slightly steeper
than 1, while, as mentioned above, the data indicate a slope of
1.45$\pm$0.04 (Boesgaard et al. 1999b). Indeed, the simplest
evolutionary considerations for both the CRS and LECR models would
predict that log(Be/H) vs. [O/H] should have a slope of 1. But, as
shown above, the delay introduced by cosmic-ray transport with
n$_{\rm H}$=0.1 will steepen the slope. However, the LECRs slow down
much faster than the higher energy CRS cosmic rays. Thus, the
calculation of Figure~5 assigns no delay to the Be deposition,
leading to a possible inconsistency between the predictions of the
CRI+LECR model and the log(Be/H) vs. [O/H] data.

\begin{figure}
\plottwo{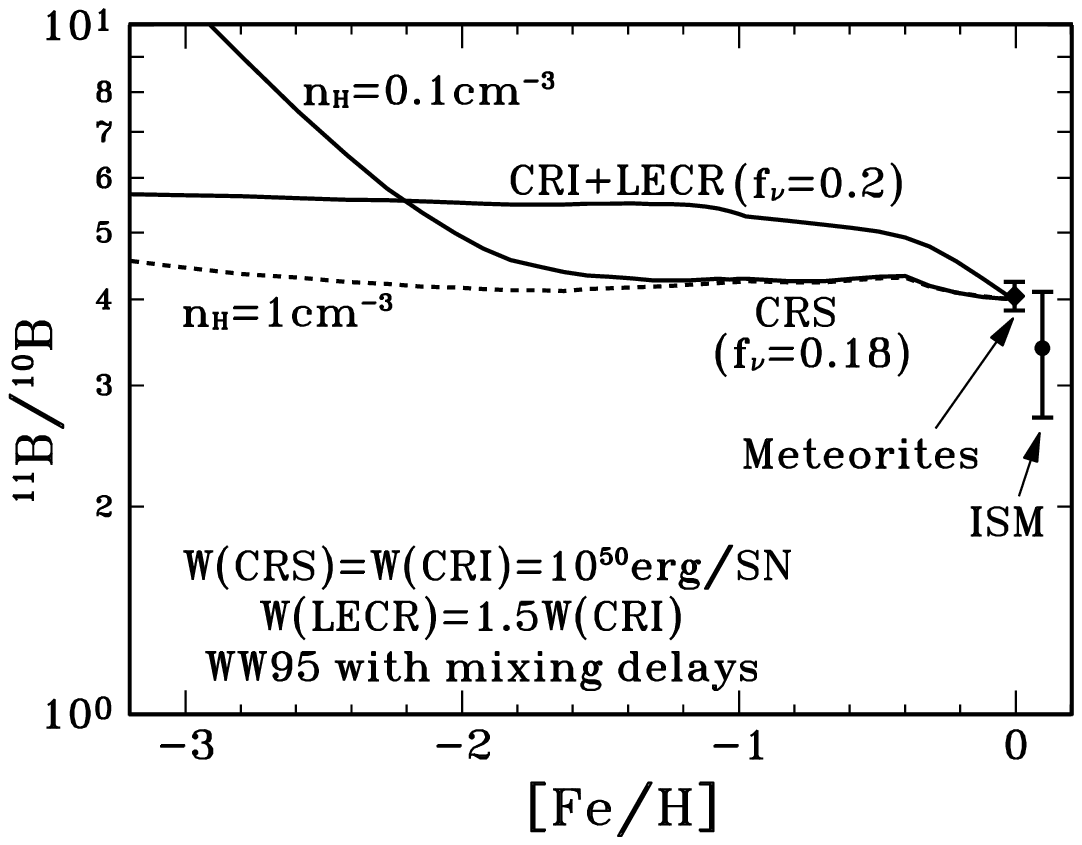}{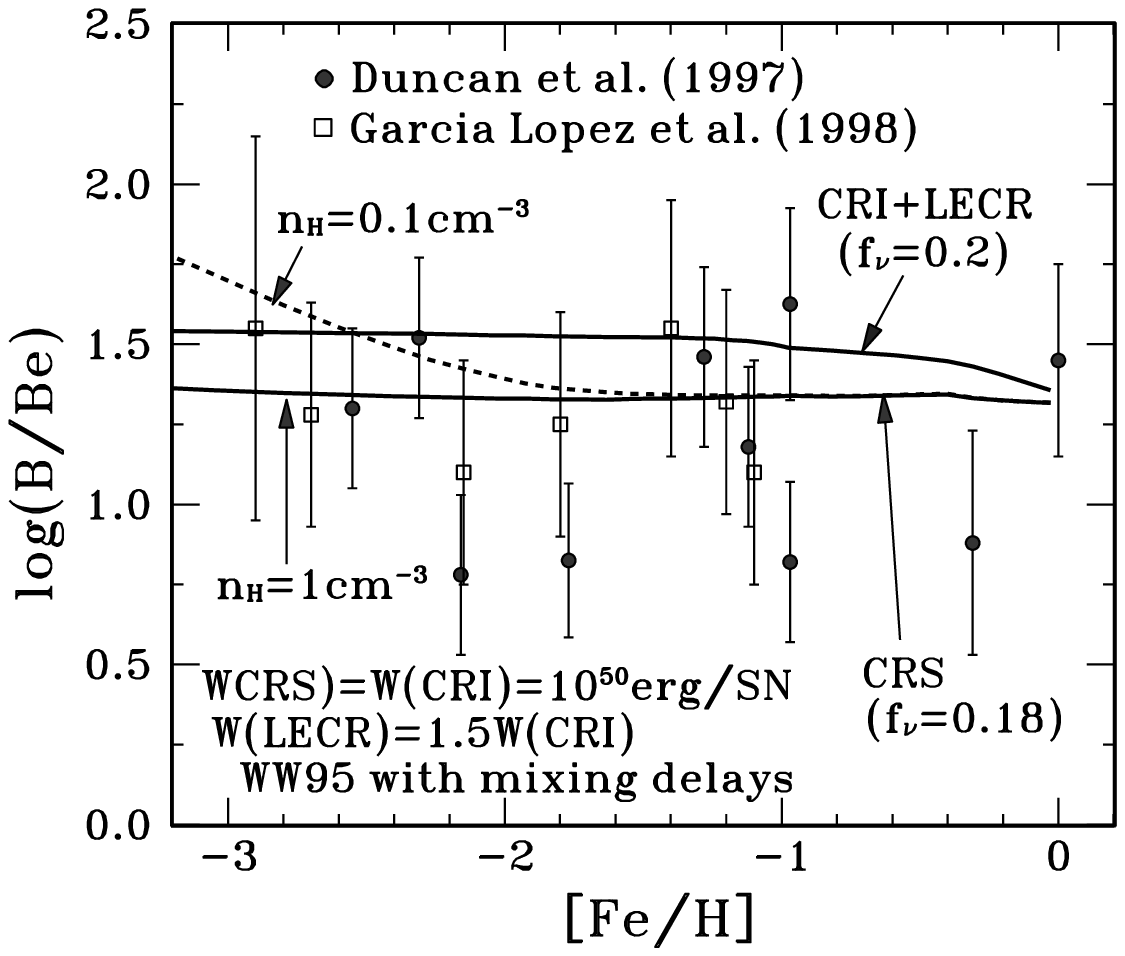} \caption{$^{11}$B-to-$^{10}$B
(left panel) and  B-to-Be (right panel) abundance ratios vs.
[Fe/H] for the CRS and CRI+LECR models. $f_\nu$ is the ratio of
the employed $\nu$-produced $^{11}$B to the nominal Woosley \&
Weaver (1995) yields. n$_{\rm H}$ is defined in the caption of
Figure~4. The B isotope data for meteorites and the ISM are from
Chaussidon \& Robert (1995) and Lambert et al. (1998),
respectively.}
\end{figure}

Figure~6 shows the boron isotopic ratio (left panel) and log(B/Be)
(right panel) vs. {Fe/H} for the CRS and CRI+LECR models. In the
evolutionary calculations, the production ratios $Q(^{11}{\rm
B}/^{10}{\rm B})=2.4$ and $Q({\rm B}/{\rm Be}) =14$ for the CRS
and CRI cosmic rays, and $Q(^{11}{\rm B}/^{10}{\rm B})=3.3$ and
$Q({\rm B}/{\rm Be}) = 22$ for the LECRs, were taken as
independent of [Fe/H]. These numerical values are from Ramaty et
al. (1997) and are valid for high energy CRS and CRI cosmic rays
and LECRs with $E_0=30$ MeV/nucleon. In order to reproduce the
meteoritic $^{11}$B/$^{10}$B both the CRS and CRI+LECR models
require the addition of $\nu$-produced $^{11}$B (Woosley \& Weaver
1995). We take into account the metallicity dependence of this
$^{11}$B production, and the fact that only the core collapse
supernovae produce $^{11}$B by neutrinos. We find that for the CRS
model the meteoritic data can be fit with $f_{\nu}$=0.18, a lower
value than we found in Ramaty et al. (2000) because of the lower
cosmic-ray energy per supernova that we use here (10$^{50}$ vs.
1.5$\times$10$^{50}$ erg that we used in that paper). The rise in
$^{11}$B/$^{10}$B and B/Be for n$_{\rm H}$ = 0.1 below [Fe/H] of
about $-2$ in the CRS model is due to the delayed deposition of
the cosmic-ray produced Be and B relative to the $\nu$-produced
$^{11}$B for which we took a short delay time, the same as for O
(1 Myr). For the CRI+LECR model, in which the LECRs employ a
larger energy per supernova, $f_\nu$ = 0.2 is required to fit the
boron isotope data. The rise in $^{11}$B/$^{10}$B and B/Be with
decreasing [Fe/H] is mostly due to the larger value of the
corresponding LECR production ratios. It is evident from Figure~6
that future high precision measurements of $^{11}$B/$^{10}$B and
B/Be as functions of [Fe/H] will distinguish between the models.

\begin{figure}
\plottwo{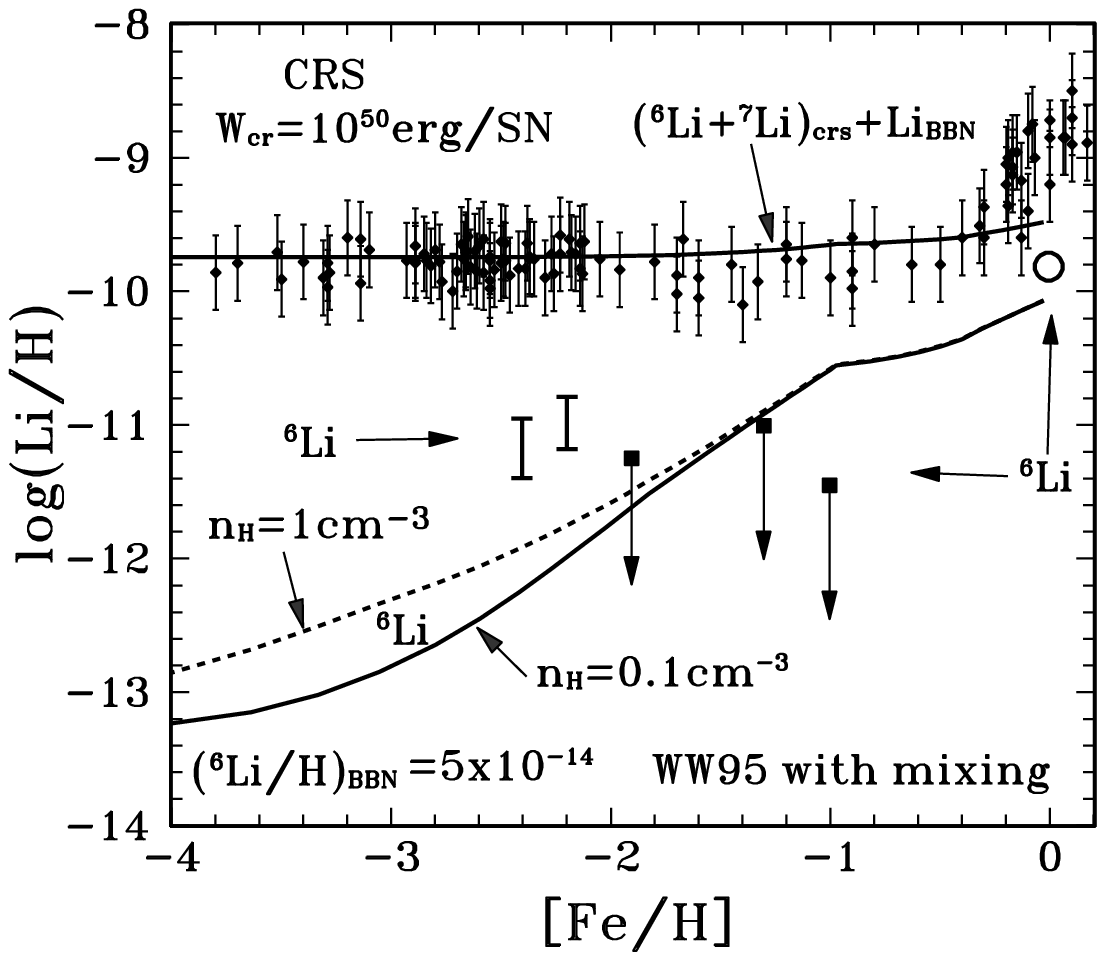}{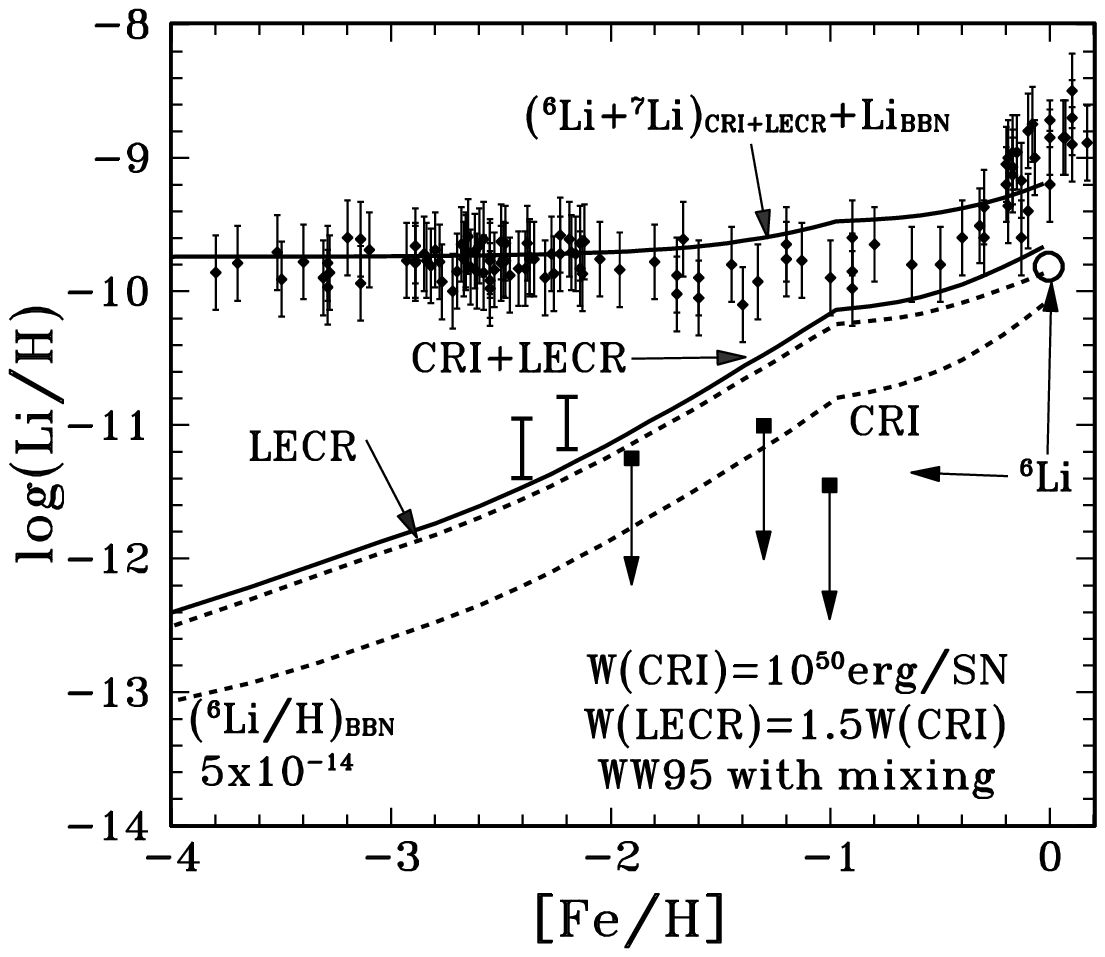} \caption{ The evolution of the
total Li and $^6$Li for the CRS (left panel) and the CRI+LECR
models (right panel). The $^6$Li data at [Fe/H] of $-$2.4, $-$2.2,
$-$1.9, $-$1.3 and $-$1 are from the summary of Hobbs (1999), and
the meteoritic value at [Fe/H]=0 is from Grevesse et al. (1996).
The total Li data is from a compilation by M. Lemoine (private
communication 1997). n$_{\rm H}$ is defined in the caption of
Figure~4. We took the primordial $^7$Li/H=1.8$\times$10$^{-10}$
(Molaro 1999).}
\end{figure}

Figure~7 shows the Li evolution. The CRS model underproduces the
$^6$Li abundance for [Fe/H]$<-2$, suggesting the existence of
pregalactic or extragalactic $^6$Li sources. With the lower
cosmic-ray normalization mentioned above, the CRS model also
slightly underproduces the meteoritic $^6$Li. With the discovery
of solar flare produced $^6$Li in the solar wind (via measurements
in lunar soil, Chaussidon \& Robert 1999), the possibility of some
locally produced $^6$Li in the solar system must be considered.
The CRI+LECR model produces more $^6$Li relative to Be, simply
because the $\alpha$$\alpha$ cross section for $^6$Li production
peaks in the nonrelativistic region.

\section{Discussion and Conclusions}

We have summarized a complete set of O, Fe and LiBeB evolutionary
calculation. We have considered the three principal evolutionary
models, CRI in which the cosmic-ray source composition at all epochs
of Galactic evolution is similar to that of the average ISM at that
epoch, CRI+LECR in which metal enriched low energy cosmic rays
(LECRs) are superimposed onto the CRI cosmic rays, and CRS in which
the cosmic-ray source, accelerated in superbubbles, has a constant
composition, independent of the ISM metallicity. By considering the
evolutionary trend of log(Be/H) vs. both [Fe/H] and [O/H], we
demonstrated that the CRI model is energetically untenable. Although
the CRI+LECR mix considered here is consistent with the Be evolution
vs. Fe, a plausible scenario for producing the required mix has yet
to be proposed.

For Fe, our code allows for a delay between nucleosynthesis and
deposition into star forming regions due to the incorporation of
the synthesized Fe into high velocity dust. This delay could
provide an explanation for the possible rise of [O/Fe] with
decreasing [Fe/H] indicated by some of the data. A test for this
scenario would be the demonstration that the abundances of
refractory $\alpha$-elements Mg, Si, Ca and Ti relative to Fe do
not increase with decreasing [Fe/H] below [Fe/H]$= -1$, but that
volatile sulfur does rise.

For the LiBeB there is also a delay. Due to the transport of the
cosmic rays, LiBeB synthesis lags behind the explosion of the
supernova responsible for accelerating the cosmic rays by as much
a hundred million years, depending on the average gas density in
the halo of the early Galaxy. We show that this delay, combined
with the delayed Fe deposition, could provide an explanation for
the steeper evolutionary trend of log(Be/H) vs. [O/H] than vs.
[Fe/H], as indicated by the recent data of Boesgaard et al.
(1999b).

The trends of $^{11}$B/$^{10}$B and B/Be vs. [Fe/H] show structure
resulting from the above mentioned delays, as well as from the
hybrid nature of the CRI+LECR model. Future observations of these
ratios may distinguish between the models. The fact that the
$^6$Li abundances at [Fe/H] below $-2$ seem inconsistent with all
the models suggests the existence of pregalactic or extragalactic
$^6$Li sources.


\begin{references}

\reference Bloemen, H. et al. 1994, \aap, 281, L5

\reference Bloemen, H. et al. 1999, \apj, 521, L137

\reference Boesgaard, A., King, J., Deliyannis, C., \&
Vogt, S. 1999a, \aj, 117, 492

\reference Boesgaard, A., Deliyannis, C., King, J., Ryan,
S., Vogt, S., \& Beers, T. 1999b, \aj, 117, 1548

\reference Bykov, A., \& Bloemen, H. 1994, \aap, 283, L1

\reference Bykov, A. M., \& Fleishman, G. D. 1992, \mnras, 15, 269

\reference Cass\'e, M., Lehoucq, R., \& Vangioni-Flam, E. 1995,
Nature, 373, 318

\reference Chaussidon, M., \& Robert, F. 1995, Nature, 374, 337

\reference Chaussidon, M., \& Robert, F. 1999, Nature, 402, 270

\reference Duncan, D. K., Lambert, D. L., \& Lemke, M., 1992,
\apj, 401, 584

\reference Duncan, D. K. et al. 1997, \apj, 488, 338

\reference Ellison, D. C., Drury, L.O'C., \& Meyer, J-P. 1997,
\apj, 487, 197

\reference Fields, B. D., \& Olive, K. A. 1999, \apj, 516, 797

\reference Fulbright, J. P, \& Kraft, R. P. 1999, \aj, 118, 527

\reference Garcia Lopez, R. J. et al. 1998, \apj, 500, 241

\reference Grevesse, N., Noels, A., \& Sauval, A. J. 1996, in:
Cosmic Abundances, eds. S. S. Holt, \& G. Sonneborn, ASP Conf.
Ser., 99, (San Francisco: ASP), 117

\reference Higdon, H. C., Lingenfelter, R. E., \& Ramaty, R.,
1998, \apj, 509, L33

\reference Higdon, H. C., Lingenfelter, R. E., \& Ramaty, R.,
1999, 26th Internat. Cosmic Ray Conf., eds. D. Kieda et al. (Salt
Lake City), 4, 144

\reference Hobbs, L. M. 1999, in: LiBeB, Cosmic Rays, and Related
X- and Gamma-Rays, eds. R. Ramaty et al., ASP Conf. Ser., 171,
(San Francisco: ASP), 23

\reference Israelian, G., Garcia Lopez, R. J., \& Rebolo, R. 1998,
ApJ, 507, 805

\reference Lambert, D. L. et al. 1998, \apj, 494, 614

\reference Lingenfelter, R. E., \& Ramaty, R., 1999, 26th
Internat. Cosmic Ray Conf., eds. D. Kieda et al. (Salt Lake City),
4, 148

\reference Lingenfelter, R. E., Ramaty, R., \& Kozlovsky, B. 1998,
\apj, 500, L153

\reference Meyer, J-P., Drury, L. O'C., \& Ellison, D. C. 1997,
\apj, 487, 182

\reference Molaro, P. 1999, in: LiBeB, Cosmic Rays, and Related X-
and Gamma-Rays, eds. R. Ramaty et al. , ASP Conf. Ser., 171, (San
Francisco: ASP), 6

\reference Parizot, E. M. G., Cass\'e, M., \& Vangioni-Flam, E.
1997, \aap, 328,107

\reference Ramaty, R. 1996, \aap (Suppl.), 120, C373

\reference Ramaty, R., Kozlovsky, B., \& Lingenfelter, R. E. 1996,
ApJ, 456, 525

\reference Ramaty, R., Kozlovsky, B., Lingenfelter, R., \& Reeves,
H. 1997, \apj, 488, 730

\reference Ramaty, R., Scully, S., Lingenfelter, R., \& Kozlovsky,
B. 2000, \apj, in press

\reference Reeves, H., Fowler, W. A., \& Hoyle, F. 1970, Nature,
226, 727

\reference Ryan, S. G., Norris, J. E., \& Beers, T. C. 1996, \apj,
471, 254

\reference Ryan, S., Bessell, M., Sutherland, R., \& Norris, J.
1990, \apj, 348, L57

\reference Timmes, F. X., Woosley, S. E., \& Weaver, T. A., 1995.
\apj(suppl), 98, 617

\reference Tsujimoto, T., \& Shigeyama, T. 1998, \apj, 508, L151

\reference Vangioni-Flam E., Cass\'e, M., Audouze, J., \& Oberto,
Y. 1990, \apj, 364, 568

\reference Vangioni-Flam, E., Cass\'e, M., Fields, B., \& Olive,
K. 1996, A\&A, 468, 199

\reference Westin, J. et al. 1999, \apj, in press
(astro-ph/9910376)

\reference Wiedenbeck, M. E. et al. 1999, \apj (Letters), 523, L61

\reference Woosley, S. E. \& Weaver, T. A. 1995, ApJS, 101, 181

\end{references}
\end{document}